# Active Microrheology of Networks Composed of Semiflexible Polymers

## II. Theory and comparison with simulations

N. Ter-Oganessian\*, D. Pink♦ and A. Boulbitch\*

\*Dept. for Biophysics E22, Technical University Munich, James-Franck-Str. 1, D-85747 Garching, Germany

♦ Phys.-Dept. St Francis Xavier University, Antigonish, Nova Scotia, Canada B2G2W5

## Abstract

Building on the results of our computer simulation[1] we develop a theoretical description of the motion of a bead, embedded in a network of semiflexible polymers, and responding to an applied force. The theory reveals the existence of an osmotic restoring force, generated by the piling up of filaments in front of the moving bead and first deduced through computer simulations. The theory predicts that the bead displacement scales like $x \sim t^\alpha$ with time, with $\alpha = 1/2$ in an intermediate- and $\alpha = 1$ in a long-time regime. It also predicts that the compliance varies with concentration like $c^{-4/3}$ in agreement with experiment.

## 1. Introduction

Actin networks play a key role in the mechanical stability of cells and for numerous mechanochemical processes[2] such as cell locomotion on surfaces[3] and the growth of cellular protrusions[4]. The viscoelastic properties of actin networks have been extensively studied by torsional rheometry[5-8], passive one-[9, 10] and two-bead[11] microrheology and force-based (oscillatory) microrheometry[12, 13].

Recently new results using pulsed force magnetic bead microrheometry were reported[14, 15] in which three regimes of bead movement were observed, each being described by the power law

$$x \sim t^\alpha \qquad (1)$$

where $x$ is the bead displacement and $t$ is the time: the short-time regime with the exponent $\alpha_1 \approx 0.75$ observed for $\tau_i \leq t \leq \tau_1$ in which the bead undergoes a displacement less than the mesh size; the intermediate-time regime characterized by $\alpha_2 \approx 0.5$, for $\tau_1 \leq t \leq \tau_2$, and the long-time viscous-like regime, for $t > \tau_2$, with the exponent $\alpha_3 \approx 1$. Here $\tau_i \approx 0.6$ms was the time resolution, the cross-over time $\tau_1$ varied between 0.03 and 0.3s while $\tau_2$ varied between 10 and 30s depending on the force applied to the bead. In the intermediate and long-time regimes the dependence of bead displacement on the concentration of actin filaments, $c$, behaved like $x \sim c^{-\gamma}$ with $\gamma \approx 1.1 \pm 0.3$ in the intermediate- and $\gamma \approx 1.3 \pm 0.3$ in the long-time regime[14].

Actin networks and constrained actin filaments were studied theoretically in a series of papers[5, 6, 16-31] and theoretical analyses of microrheology have recently been reported in[32-34]. Up to now no theoretical explanation of the $x \sim t^{1/2}$ power law of the bead motion in active microrheology of semiflexible entangled networks has been proposed.

In the preceding paper[1] we reported the results of computer simulation of magnetic bead microrheometry to study a network of semiflexible polymers modeling entangled actin filaments in an aqueous solution. The simulations revealed two regimes of bead motion. The well-known initial regime[9, 11, 21, 23-25] was characterized by the exponent $\alpha_1 \approx 0.75$. Displacement of the bead in this regime was smaller than the network mesh size in accord with the observations[14]. It was followed by the regime with $\alpha_2 \approx 0.5$ in which the bead displacement took the form:

$$x \sim D_\parallel^{1/2} f t^{\alpha_2} / c^{\gamma_2} \qquad (2)$$

with the exponents $\alpha_2 \approx 0.5$ and $\gamma_2 \approx 1.4$. Here $D_\parallel$ is the longitudinal diffusion coefficient. A total displacement of several mesh sizes was achieved by the end of simulation. These two regimes observed in the simulations correspond to the short- and intermediate- time regimes (above)[14, 15] and are characterized by the same exponents.

The simulations showed that, during bead motion in the intermediate-time regime, the polymers piled up in front of the bead, while the region behind the bead was almost free of polymers. Further, the simulations demonstrated that the resistance of the network to bead motion is due mainly to the steric repulsion of these piled up polymers. Finally, the simulations[1] showed that the polymers in front of the bead take part, on average, in a diffusive motion in the direction of the bead motion. The diffusion coefficient of this motion is close to the longitudinal diffusion coefficient describing the free diffusion of polymers in the bulk[1].

Based on these findings, this paper formulates an analytical approach deriving the intermediate- and the long-time regimes and describing both the experimental observations[14, 15] and the results of our simulations[1].

In Section 2 we deduce the force resisting the bead motion on the assumption that a clump of polymers has been formed in front of the moving bead. In Section 3 we describe the formation of the clump and obtain equations of motion of the bead in the intermediate-time regime. In Section 4 we extend our approach to describe the long-time regime of bead motion.

In Section 5 we compare our predictions with measurements and simulations, and discuss the relation of our findings to previous experimental and theoretical results. Section 6 summarizes our results.

## 2. Osmotic force

Actin networks are well described by a reptation-tube model[26, 27]. Within this approach each filament is considered to be confined to a tube with diameter equal to the mesh size $\sim \xi$ accounting for the steric contribution of the surrounding filaments. This yields the free energy per filament[17, 19, 20, 35]

$$F_f \approx \kappa k_B T L / L_e \qquad (3)$$

where $\kappa \approx 2.46$ is a geometric factor, $L_e \sim L_p^{1/3} \xi^{2/3}$ is the entanglement length[16] and $L_p$ is the persistent length of the filament. The term originating from the translational entropy of the filaments is omitted in (3), since its effect on the resistive force is small.

In a system with $N$ polymers of length $L$ homogeneously distributed in volume $V$ the mesh size of the network, $\xi$, can be estimated according to the relation $V = gNL\xi^2$, where $g$ is a geometric factor. The concentration of the polymers is defined as $c = N/V$ yielding

$$c = 1/gL\xi^2 \qquad (4)$$

In the following it is convenient to write all expressions in terms of the concentration $c^*$ of segments of filaments with length $L_e$ according to the definition $c^* = cL/L_e$:

$$c^* = 1/gL_p^{1/3}\xi^{8/3} \qquad (5)$$

In the following these segments play an important role and therefore, for the sake of brevity, we introduce a special term, "entanglement segments", for them. According to (3) each entanglement segment carries the thermal energy $\kappa k_B T$.

The free energy of the network takes the form $F = NF_f$:

$$F = \kappa k_B T \left[ \frac{g(LN)^4}{L_p V} \right]^{1/3} \tag{6}$$

Making use of Eq. (6) one calculates the pressure $p = -(\partial F / \partial V)_T$ and, using Eq. (5) one finds:

$$p = \frac{\kappa}{3} c^* k_B T \tag{7}$$

The pressure (7) depends on the mesh size like $p \sim \xi^{-8/3}$. It arises due to the decrease of entropy of a filament subjected to a tube. It can be considered as the osmotic pressure of the entanglement segments.

The beads used in the measurements[14, 15] and modeled in the simulations[1] are about 5 to 10 times larger than the mesh sizes of the networks, and cannot squeeze through the network. For this reason during its motion, the bead compresses the network ahead of it, thus increasing the local concentration of filaments (Fig. 1). The mesh size in front of the bead, $\xi = \xi_f$, (Fig. 1 b) thus, becomes smaller than the mesh size, $\xi = \xi_0$, far from the bead (Fig. 1 a).

It has been shown by the simulations that the main contribution to the resistance is due to the steric repulsion between the bead and filaments which gives rise to the pressure $p$ exerted by the filaments on the bead surface[1]. Further, the simulations show that the filaments are piled up in front of the bead, while behind the bead a region almost free of filaments develops during its motion[1] and hence, it is a good approximation that essentially zero pressure is exerted by filaments on the back hemisphere. Thus, the force $f \sim \pi R^2 p$, where $R$ is the bead radius, acts on the front surface of the bead and resists the bead motion. One finds

$$f \sim \frac{\kappa}{3} \pi R^2 c_f^* k_B T \tag{8}$$

where $c_f^* = 1/g L_p^{1/3} \xi_f^{8/3}$ is the concentration of the entanglement segments in the neighborhood of the front of the bead where the mesh size $\xi$ is equal to $\xi_f$. The force (8) represents the steric repulsion of the bead by the polymers piled up front of it. It is related to the osmotic pressure of the entanglement segments and we refer to it as the "osmotic force".

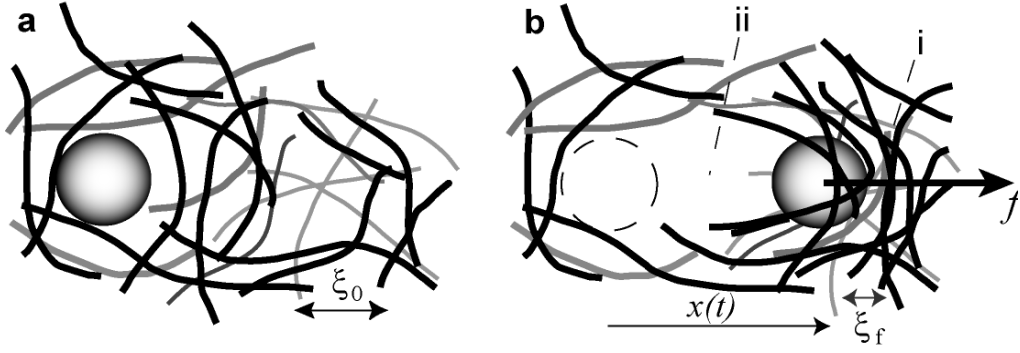

Fig. 1. Schematic view of the bead embedded in a network of semiflexible filaments. (a) Bead in network at rest. (b) During its motion the bead piles up the filaments ahead of it (i), while the region behind the bead (ii) contains fewer filaments. The dashed circle shows the initial position of the bead, $\xi_f$ indicates the mesh size in the vicinity of the bead, and $\xi_0$ is the mesh size far from the bead.

### 3. Equation of motion of the bead: the intermediate-time regime

In the beginning of the motion, when $t < \tau_1$ and $x < \xi$, the bead response is due to a dynamic bending of a few filaments and obeys the law[21] $x \sim t^{3/4}$ (see Appendix A).

At $t > \tau_1$ the short-time regime is followed by the intermediate-time regime in which the bead motion depends upon the response of a large number of those filaments ahead of the bead and crossing the bead's path. The main contribution to the resistance force is described by (8). The concentration of entanglement segments, $c_f^*$, entering Eq. (8) is to be calculated directly in front of the bead and should be related to the concentration of segments $c_0^*$ (or to the filament concentration $c_0$) far from the bead. The latter values are set by the preparation of the network and we assume them to be known. Basing on the simulations[1] we also assume that the entanglement segments move by diffusion with the diffusion coefficient $D$.

Consider a bead moving over a distance $x(t)$ during the time $t$. The bead compresses $n_+$ entanglement segments ahead of it given by $n_+ \sim c_0^* \pi R^2 x(t)$ where $c_0^* = g^{-1} L_p^{-1/3} \xi_0^{-8/3}$ is the concentration of the entanglement segments far from the bead, where the mesh size is $\xi_0$. These entanglement segments however, redistribute themselves by diffusing over a typical distance $\sim (Dt)^{1/2}$. If $\tau_1 < t < \tau_2$ where

$$\tau_2 = R^2 / 2D \tag{9}$$

there is not enough time for the entanglement segments to diffuse over the distance $\sim R$ away from the bead's path. Accordingly, for $t < \tau_2$ one needs consider only the diffusion of entanglement segments in the $Ox$ direction. This motion results in a clump of filaments of thickness $\sim (Dt)^{1/2}$ formed in front of the bead so that the entanglement segments redistribute themselves over the volume $\delta V \sim \pi R^2 (Dt)^{1/2}$. Accordingly, the concentration, $c^* \sim n_+ / \delta V$, in the clump in front of the bead is,

$$c^* \sim \frac{x(t)}{(Dt)^{1/2}} c_0^* \tag{10}$$

Substitution of (10) into the resistive force (8) yields the equation describing the bead motion during the time interval $\tau_1 < t < \tau_2$:

$$x \sim \frac{3 g L_p^{1/3} \xi_0^{8/3} D^{1/2}}{\kappa \pi R^2 k_B T} f t^{1/2} \tag{11}$$

### 4. The long-time regime of bead motion

If $t > \tau_2$ the polymers have enough time to diffuse sideward and to escape from the path of the bead. In this case the motion of the bead becomes a steady-state motion as soon as the number of the entanglement segments, $n_+$, being captured by the clump in front of the bead becomes equal to the number $n_-$ diffusing sidewise away from the bead's path. As soon as this regime is achieved the bead moves with a constant velocity $v$. The number of entanglement segments $n_+$ compressed by the bead within the time $t$ can be estimated as

$$n_+ \sim c_0^* \pi R^2 v t \qquad (12)$$

The flux of entanglement segments is $\mathbf{j} = -D\nabla c^*$. In this regime the only relevant length scale is the bead size $\sim R$. Accordingly the size of the clump, $l$, both along and perpendicular the $Ox$ direction can be estimated as $l \sim R$. One finds an estimate of the flux of segments diffusing away from the clump to be of the form $j \sim -Dc_f^*/R$. The number of entanglement segments diffusing sidewise away from the bead's path is

$$n_- \sim 2\pi R l j t \qquad (13)$$

Applying the condition of steady-state motion, $n_+ = n_-$, one finds the concentration of entanglement segments in the clump, $c_f^*$, to be

$$c_f^* \sim R c_0^* v / 2D \qquad (14)$$

Substitution of the concentration (14) into the expression for the resistive force (8) yields the equation for steady-state motion,

$$x \sim \frac{6 g L_p^{1/3} \xi_0^{8/3} D}{\kappa \pi R^3 k_B T} f t \qquad (15)$$

## 5. Discussion

### 5.1. Diffusion coefficient of entanglement segments

Results of our simulation study[1] indicate that the dilatation mode of motion of the clump in front of the bead is diffusive and its diffusion coefficient, $D$, is close to that of the longitudinal diffusion of filaments, $D_\parallel$. On the other hand, the transverse motion of the filaments in front of the bead leading to a compression or dilatation of this region is more plausible. The transverse motion differs considerably form the longitudinal diffusion[1, 36].

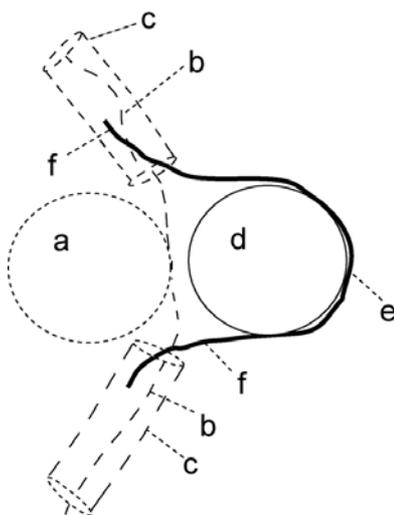

Fig. 2. Enforced motion of a filament crossing the bead's path. An initial position of the bead is shown by (a). (b) shows tails of one of the filaments which crosses the bead's path in front of it. Other such filaments are not shown. The network (not shown) gives rise to the reptation tubes (c) in which the tails move. Displacement of the bead to a new position (indicated by d) requires the movement of those segments crossing the bead's path (e). This requires that the tails slip out of their tubes, thus taking a new configuration (f).

This seeming contradiction can be explained as follows. A segment of a filament which crosses the bead's path has the length of about 5μm. This is smaller than the filament contour length (≈20μm). For this reason a clump of polymers in front of the bead is connected by filament tails with the rest of the network, these tails being longer than the size of the clump (Fig. 2). When the bead compresses the clump during its motion, the segments of those filaments in the clump undergo mainly a transverse motion. However, in order to enable such a motion, a longitudinal displacement of the tails connecting the clump with the network along their reptation tubes is required. This is shown schematically in Fig. 2 (b) and (f). Since the lengths of the tails are larger then the clump size, the clump compression and dilatation by the bead is dominated by the longitudinal diffusion of filaments. For this reason the diffusion coefficient describing the dilatation dynamics of filaments in front of the bead is close to the longitudinal diffusion coefficient of filaments $D \approx D_\parallel$. This conclusion agrees with results of the simulation[1].

## 5.2. Comparison with experiments

The creep compliance of the bead, $J(t)$, defined by

$$J(t) = 6\pi R x(t)/f \qquad (16)$$

was measured in the experiments reported in[14, 15]. It was found that, in each of the three regimes, the compliance could be described by the power law

$$J(t) = A_i t^{\alpha_i} + B_i \qquad (17)$$

where $i = 1, 2, 3$ denotes the short-time, intermediate and long-time regimes characterized by the exponents $\alpha_1^{(exp)} \approx 0.75$, $\alpha_2^{(exp)} \approx 0.5$ and $\alpha_3^{(exp)} \approx 1.0$, and amplitudes $A_i$ and offsets $B_i$ with $B_1=B_2=0$[14, 15].

In Section 3 (above) and in Appendix A these regimes are described analytically. Our estimates on the short-time regime (Eq. (A4) Appendix A) reproduces the well-known[21] exponent $\alpha_1^{(th)} = 3/4$. In the long-time regime we obtained the expected exponent $\alpha_3^{(th)} = 1$ (Eq. (15)) describing the viscous-like motion of the bead. Finally in the intermediate-time regime we derived the exponent $\alpha_2^{(th)} = 1/2$ (Eq. (11)) in agreement with the observations reported in[14, 15] as well as our simulations[1]. The theory presented here predicts the pre-factors $A_i$ to take the forms:

$$A_1 \approx \frac{R}{32} \left[ \frac{\pi}{k_B T L_p (4\eta)^3} \right]^{1/4} \qquad (18)$$

$$A_2 \sim \frac{18 g L_p^{1/3} \xi_0^{8/3} D^{1/2}}{\kappa R k_B T} \qquad (19)$$

$$A_3 \sim \frac{36gL_p^{1/3}\xi_0^{8/3}D}{\kappa R^2 k_B T} \tag{20}$$

where $\eta$ is the water viscosity.

Since, according to (4), the concentration of filaments far from the bead, $c_0$, scales like $c_0 \sim \xi_0^{-2}$, Eq. (19) and (20) yield $A_{2,3} \sim c^{-\gamma_{2,3}}$. Our approach predicts $\gamma_{2,3}^{(th)} = 4/3$ in agreement with the experimentally measured value, $\gamma_2^{(exp)} \approx 1.1 \pm 0.3$ and $\gamma_3^{(exp)} \approx 1.4 \pm 0.3$ [14], as well as with the value $\gamma^{(sim)} \approx 1.4$ found in the simulations[1].

In order to make estimates we used $k_B T \approx 4 \times 10^{-21} \text{J}$, $\eta \approx 10^{-3} \text{Pa s}$, the values $\xi_0 \approx 0.3 \mu\text{m}$, $L \approx 20 \mu\text{m}$, $R \approx 2.25 \mu\text{m}$ and $L_p \approx 17 \mu\text{m}$, obtained from the measurements[14, 15]. In addition, from experiment[37] $D \sim 10^{-13} \text{m}^2/\text{s}$. If the filaments of the network lie along a primitive cubic lattice, one would find $g = 1/3$ which we will use in making estimates. This yields $A_1^{(th)} \approx 11.5 \text{Pa}^{-1}\text{s}^{-3/4}$, $A_2^{(th)} \sim 10 \text{ Pa}^{-1}\text{s}^{-1/2}$ and $A_3^{(th)} \sim 1 \text{ Pa}^{-1}\text{s}^{-1}$ in good agreement with the experimental values $A_1^{(exp)} \approx 18.9 \text{Pa}^{-1}\text{s}^{-3/4}$, $A_2^{(exp)} \approx 11.2 \text{Pa}^{-1}\text{s}^{-1/2}$ and $A_3^{(exp)} \approx 2.3 \text{ Pa}^{-1}\text{s}^{-1}$ reported in[14].

It is generally expected that in the long-time regime the bead motion is dominated by viscosity yielding $x = ft/6\pi R\eta_{net}$ and thus,

$$A_3 = \eta_{net}^{-1} \tag{21}$$

where $\eta_{net}$ is the (effective) viscosity of the network. The bead displacement (15) indeed depends linearly upon time. However, in the derivation of (15) only the osmotic mechanism was accounted for, while the viscosity was neglected. Using the above estimates one finds the effective viscosity of the network in the long-time regime $\eta_{net} = 1/A_3^{(th)} \sim 1 \text{Pa s}$ to be in agreement with the earlier measurements[38]. The good agreement between our estimate of $A_3$ (above) with the measured value shows that the osmotic mechanism indeed makes the main contribution to the network resistance during the long-time regime. The bead motion in this regime is thus, viscous-like, rather than viscous.

Equating the bead compliance in the short- and the intermediate-time regimes $A_1 \tau_1^{3/4} = A_2 \tau_1^{1/2}$ and using (18, 19) one finds the crossover time $\tau_1$:

$$\tau_1 = \frac{q L_p^{7/3} \xi_0^{32/3} D^2}{R^8} \left( \frac{\eta}{k_B T} \right)^3 \qquad (21)$$

where the numerical factor is $q = 4^3 \times 192^4 / \pi \kappa^4 \approx 7.6 \times 10^8$. One then finds $\tau_1 \approx 0.35 \text{s}$ in agreement with the observations[14, 15].

Analogously, by equating the compliance of the intermediate- to that of the long-time regime $A_2 \tau_2^{1/2} = A_3 \tau_2$ Eq. (19), (20) one finds the expression (9) for the characteristic time $\tau_2$. Estimates yield $\tau_2^{(th)} \sim 10 \text{s}$, which is of the order of magnitude of that observed to vary between 8 and 30s[14].

## 5.2. Comparison with results of our computer simulations

Our computer simulation[1] of the enforced motion of a bead through an entangled solution of the semiflexible polymers utilized Dissipative Particle Dynamics[39]. The method used 24 phenomenological parameters describing the geometry and the interactions of the objects whose values were in accord with the choices of others[1]. However, in such a case one cannot expect to compare numerical data obtained in simulations with those from experiments. On the other hand the power laws found by the simulations can be compared with those obtained in experiments and with predictions of the analytical theory presented here. In the intermediate regime $i = 2$ the relation (2) was found for the bead displacement by the simulations (Eqs. (1) and (4) in that paper[1]) in agreement with the predictions (11) of the analytical theory.

In order to compare the analytical predictions (19) with the value of the pre-factor $A_2$ obtained by the simulation, we substituted the values $\xi = 1.57$, $R = 10$, $L_p = 150$ length units, $k_B T = 1$ and $D = 0.032$ from the simulations into Eq. (19). We found that $A_2 \approx 0.76$ is in a good agreement with the value $A_2^{(sim)} \approx 0.32$ obtained by the computer simulations.

Both simulations and theory predict a linear dependence of the bead displacement on the applied force. In experiments however, a weak non-linearity was observed[14, 15]. This indicates that the experimental system possesses interactions, beyond the steric and viscous interactions included in those models, which are not accounted for either by the theory or by the model used in the computer simulation. This non-linearity observed in the force-dependence of $A_2$ may be due to cross linkers remaining in solvent, or divalent ions, always present in the solvents for actin, and which could act as weak cross-linkers.

## 5.3. An $x \sim t^{1/2}$ regime in other systems

A binary correlation function and a mean square displacement obeying the power law $<x^2(t)> \sim t^{1/2}$ has been observed in actin networks using passive two-bead microrheometry[11]. The network concentration and the time-domain in which such a behavior was observed[11] correspond to those in experiments using active probing of actin network[14, 15]. We note that this behavior is related to the compliance depending on time like $J(t) \sim t^{1/2}$ as described in our paper in the intermediate-time regime, and observed in[14, 15]. However, a detailed theory of this phenomenon for the case of passive two-bead microrheometry is beyond the scope of the present paper.

The viscoelastic properties of complex media are often modeled in terms of equivalent mechanical circuits[40]. This approach was applied to analyze the micromechanical behavior of actin networks, the cytoplasm and cellular membranes[12, 38, 41-45]. It would be possible to fit the compliance such as that of the actin network reported in[14, 15] by an equivalent mechanical circuit. However, an accurate fitting of such a regime would require us to introduce many parameters, via springs and dash pots, which would, however, incorrectly describe the essential physics of the phenomenon. The reason that the compliance increases like $J(t) \sim t^{1/2}$ in the intermediate-time regime because of the osmotic mechanism arising from polymer compression by the moving bead and this is fundamentally different from, and cannot be reduced to the viscoelastic characteristics of the network.

In recent papers[32, 33] a phenomenological theory of microrheological measurements was formulated. It is based on a two-fluid hydrodynamic approach[46] in which the actin network was considered as an elastic medium of constant concentration viscously coupled to the penetrating water. As we already discussed at $0 < t < \tau_1$ (21) the time is not enough for osmotic pressure to contribute significantly to the bead motion and the approximation of a constant concentration used in[32, 33] is applicable. At $t > \tau_1$ the osmotic pressure becomes significant. In this case our approach may be combined with that of papers[32, 33] by accounting for the time- and coordinate-dependent concentration of filaments obeying the diffusion equation.

In general, the $x \sim t^{1/2}$ power law (or, equivalently, the power law $G_\omega \sim \omega^{1/2}$ describing the dependence of the shear modulus on frequency) indicates that the resistance mechanism is dominated by diffusion, but does not indicate the mechanism itself. The power law $G_\omega \sim \omega^{1/2}$ has been predicted theoretically for flexible polymers, for which $L/L_p \gg 1$ by accounting for the diffusion of their excess lengths along the reptation tubes[28]. In the networks studied both here and in our simulations[1], as well as in those used in measurements[14, 15], the polymers are semiflexible, $L/L_p \approx 1.2$, and, therefore, the mechanism responsible for the intermediate-time behavior observed in[14, 15] differs from that discussed in[28].

Kollmann and Nägele developed a theory describing the diffusion of a spherical macro-ion in a multicomponent colloidal dispersion. They predicted a perturbation, $\delta\zeta$, of its short-time friction coefficient, $\zeta$, by the electrolyte[47]. If the hydrodynamic interactions were "switched off" the asymptotic behavior of this perturbation would be $\delta\zeta \sim t^{-1/2}$. However, accounting for the far-field hydrodynamic interactions between the tracer and the micro-ions removes the singularity yielding $\delta\zeta$ to be regular at $t=0$ [47].

Although the entangled network of semiflexible filaments differs considerably from the multicomponent dispersion of spherical colloidal particles studied in[47], the intermediate-time regime recently observed in actin networks[11, 14, 15] and predicted by the simulations[1] is qualitatively comparable with the short-time behavior $\delta\zeta \sim t^{-1/2}$ predicted for a colloidal dispersion[47]. In the latter the hydrodynamic interactions equalize concentrations of colloidal particles in front of and behind the bead and the difference in the osmotic pressure in front of and behind the bead vanishes. In the densely entangled solution of polymers, their low mobility prevents such an equalization of the density. For this reason the osmotic force exists in the network and (as has been shown here) dominates the resistance of the gel to the bead motion giving rise to the bead motion obeying the law $x \sim t^{1/2}$.

In general, arising of the osmotic force discussed in this paper requires two conditions: (i) density inhomogeneity of the complex fluid must be formed (giving rise to the spatial inhomogeneity of the osmotic pressure) and (ii) a slow diffusive mode must be responsible for the decay of this inhomogeneity. Such conditions are often fulfilled on the mesoscale in complex fluids. In the paper[48] resistance of adhered biomembranes to enforced unbinding has been recently explained by a transient inhomogeneous osmotic pressure of mobile ligand-receptor pairs.

**Summary**

Based on our previous computer simulations[1] we have proposed a mechanism responsible for the resistance of an actin network to the enforced motion of an embedded bead. It originates in the osmotic pressure exerted on the bead surface by the actin filaments. The pressure arises, since the moving bead piles up filaments in front it, while many fewer filaments remain behind the bead. This mechanism describes the bead motion, obeying the power law $x \sim t^{1/2}$, and the subsequent viscous-like regime, $x \sim t$, as well as the dependence of the bead response on the actin concentration recently reported in[14, 15].

## Appendix A

### Approximate description of the short-time regime

The enforced bead motion obeying the power law $x \sim t^{3/4}$ has been described theoretically by a mechanism which accounts for a dynamical bending of a filament[21]. The solution was obtained in[21] in terms of a Green function. Based on the idea of that paper[21], we give here an approximate calculation of the bead displacement, which enables us to estimate the compliance in the short-time regime and to compare this prediction with our observations in Section 5.

During the short-time regime of motion the bead traverses a distance less than the mesh size ($\sim 0.1 \mu m$ [14]) and, therefore, deforms only one (or a few) filaments. If a local force $f$ is applied to the filament somewhere far from its ends at the initial time then, after a time $t$, a portion of the filament will be deformed. The length of the deformed part of the filament, $\Lambda$, can be estimated using the equation of filament motion:

$$4\pi\eta \frac{\partial x}{\partial t} = k_B T L_p \frac{\partial^4 x}{\partial s^4} + f\delta(s) \tag{A1}$$

where $4\pi\eta$ is an estimate of the friction coefficient, $\eta$ is the water viscosity, $k_B T L_p$ is the filament bending rigidity, $x$ is the filament displacement, $s$ is the coordinate along the filament and $\delta(s)$ is the $\delta$-function. This yields the time-dependence of the length of the deformed portion of a filament:

$$\Lambda = \left(\frac{k_B T L_p t}{4\pi\eta}\right)^{1/4} \tag{A2}$$

Assuming that the increase of the length $\Lambda$ is slower than the filament bending, one can describe the bending by the static equation

$$\partial^4 x / \partial s^4 = 0 \tag{A3}$$

with the boundary conditions: (i) $k_B T L_p \partial^3 x(0)/\partial s^3 = f$ describing the force applied at the point $s=0$, (ii) $\partial^2 x(0)/\partial s^2 = 0$ describing the requirement of zero torque applied to the filament by the bead and (iii) $x(\pm\Lambda) = \partial x(\pm\Lambda)/\partial s = 0$ which approximately describes the filament configuration at $l = \pm\Lambda$. Utilizing the well known solution for elastic rods[49] one finds $x = x(0) = f\Lambda^3/192 L_p k_B T$. Making use of (A2) one obtains the bead displacement

$$x(t) \sim \frac{t^{3/4}}{192\left[k_B T L_p (4\pi\eta)^3\right]^{1/4}} f \tag{A4}$$

The bead displacement (A4) represents an approximation form of the Green function obtained in[21] at $s = 0$.

**Acknowledgements**: A.B. was supported by the grant DFG (Deutsches Forschungsgemeindschaft) SA 246/28-4. This work was supported by NSERC of Canada under a Discovery Grant to D.A.P. A.B. is grateful to G. Nägele for the useful discussion.